# Electron spin-lattice relaxation in solid ethanol: Effect of nitroxyl radical hydrogen bonding and matrix disorder

Marina Kveder, Dalibor Merunka, Milan Jokić, Janja Makarević, and Boris Rakvin

*Ruđer Bošković Institute, Bijenička 54, 10000 Zagreb, Croatia*

The electron spin-lattice relaxation of 2,2,6,6-tetramethyl-1-piperidine-1-oxyl and 4-oxo-2,2,6,6-tetramethyl-1-piperidine-1-oxyl was measured at temperatures between 5 and 80 K in crystalline and glassy ethanol using *X*-band electron paramagnetic resonance spectroscopy. The experimental data at the lowest temperatures studied were explained in terms of electron-nuclear dipolar interaction between the paramagnetic center and the localized excitations, whereas at higher temperatures low-frequency vibrational modes from the host matrix and Raman processes should be considered. The strong impact of hydrogen bonding between the dopant molecule and ethanol host on the spin relaxation was observed in ethanol glass whereas in crystalline ethanol both paramagnetic guest molecules behaved similarly.

Coupling of the electron spin to disorder modes of various doped matrices has been extensively studied due to the sensitivity of the approach toward dynamical properties of the observed systems.[1–3] Research on disordered solids has shown that nitroxyl radicals can contribute toward the characterization of glass-forming materials,[4–6] providing experimental data for the development of self-consistent theories of molecular dynamics in glasses in general.[7–10] The work presented here has been in part motivated by the lack of nitroxyl spin-lattice relaxation-time data measured below 20 K in disordered solids[11,12] and, to the best of the authors' knowledge, by the very few examples comparing paramagnetic relaxation rate data in glassy and crystalline states of the same compound.[13]

Solid ethanol has been found to be a very convenient model system for the investigation of molecular solids, as it can be easily prepared in phases characterized by different types of disorder.[14,15] In our previous studies we have shown how, within the course of an *X*-band electron paramagnetic resonance (EPR) experiment, glassy and crystalline ethanol can be studied on the very same sample using incorporated nitroxyl radicals.[5] Since nitroxyl radicals can be purposely tailored, in the context of this study we have chosen two almost identical paramagnetic probes, which differ only in one carbonyl group. We focused on the influence of hydrogen bonding between the incorporated paramagnetic guest molecule and the host matrix on the microscopic nature of probe/matrix dynamics. The central point is the comparative analysis of spin relaxation in crystalline and glassy states of the same host material. The experiments were performed in the temperature range 5–80 K, which is well below the ethanol glass transition (95 K).[14]

The liquid ethanol [anhydrous, min. 99.8% (GC), p.a. from Kemika, Zagreb] and hexadeuteroethanol (deuteration degree >99.5% from Uvasol, Merck) were doped with the nitroxide paramagnetic spin-probe 2,2,6,6-tetramethyl-1-piperidine-1-oxyl (TEMPO) or 4-oxo-2,2,6,6-tetramethyl-1-piperidine-1-oxyl (TEMPONE) from Aldrich, at a concentration of 0.7 mM. Glassy and crystalline states of ethanol were prepared as described.[16] It should be stressed that these two ethanol polymorphs are the most different ones[17] and regarding EPR spectroscopy could be unambiguously and reproducibly assigned from the thermal history of the spectra.

EPR measurements were performed using a Bruker E-580 FT/CW *X*-band spectrometer equipped with an Oxford Instruments temperature unit (±0.1 K). Spin-lattice relaxation times were determined by the inversion recovery method using an echo detection sequence[18] with the pulse separation time of 200 ns and $\pi$ pulse duration of 112 ns. The measurements were performed at the central magnetic field position of the EPR spectrum. The magnetization recovery curves were fitted to the biexponential function wherein only the longer component was considered as an "effective" spin-lattice relaxation time, $T_1^*$.[19]

*EPR data.* From the apparent *X*-band cw-EPR rigid limit spectra,[16] maximal hyperfine splittings, $2A_{max}$, were estimated in both crystalline and glassy ethanol (Table I). That TEMPONE has a smaller hyperfine splitting than TEMPO is characteristic of the paramagnetic molecule itself, observed also in liquid ethanol. Differences between $2A_{max}$ for nitroxyl radicals incorporated in crystalline versus glassy ethanol are in the range of 0.2 mT, being larger in the former state of the host matrix due to the larger anisotropy of molecular packing.

Pulsed EPR experiments indicate a shorter $T_1^*$ for TEMPO than for TEMPONE in protonated ethanol glass throughout the temperature range studied (Fig. 1). The phenomenon can be ascribed to the effect of hydrogen bonding between TEMPONE and ethanol molecules since the theoretical analysis of fragile liquids experiencing high viscosity due to the hydrogen-bond network has shown that the relaxation time is expected to be longer the higher the number of hydrogen bonds.[20] In order to better characterize the coupling of the probe molecules with the glassy matrix, experiments were also performed in deuterated ethanol glass (Fig. 1). The isotope substitution induced a significant decrease in the spin-lattice relaxation rate of TEMPO, whereas it had hardly any effect on TEMPONE data, a phenomenon which will be discussed in detail in the next section. Upon the ethanol glass-crystalline transformation, the spin-lattice relaxation did not show any significant difference between TEMPO and TEMPONE incorporated in crystalline ethanol [Fig. 2(a)]. This observation points to the similar molecular packing in the vicinity of the probes due to the hydrogen bonding between the host molecules themselves.[15] The ratio of $T_1^*$ data for crystalline versus glassy ethanol is shown in Fig. 2(b). It can clearly be seen that TEMPO has a much higher sensitivity than TEMPONE to the degree of disorder in the host matrix. For instance, the former paramagnetic probe exhibits approximately three times shorter $T_1^*$ in ethanol glass than in crystalline ethanol at lowest temperatures studied. At the same time, the scattering of $T_1^*(c)/T_1^*(g)$ data for TEMPONE prevents any serious conclusion about the difference between spin-lattice relaxations in crystalline/glassy state of the host matrix. The reason for different $T_1^*(c)/T_1^*(g)$ temperature behavior between TEMPO and TEMPONE can be searched in different processes governing the spin relaxation.

TABLE I. Comparison of the apparent maximal hyperfine splittings, $2A_{max}$, at 80 K estimated from cw-EPR spectra (Ref. 16) for nitroxyl radicals in ethanol glass (g) and crystalline (c) ethanol.

| Nitroxyl radical | $2A_{max}$ (mT)$^g$ | $2A_{max}$ (mT)$^c$ |
| --- | --- | --- |
| TEMPO | 7.189 ± 0.008 | 7.44 ± 0.03 |
| TEMPONE | 6.984 ± 0.006 | 7.180 ± 0.005 |

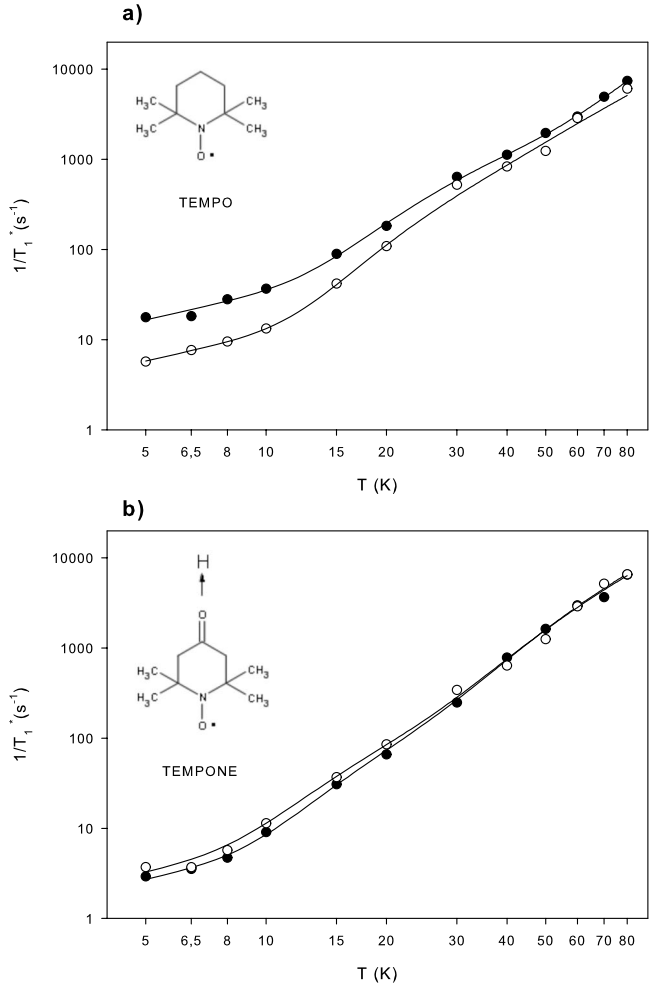

FIG. 1. Temperature dependence of the spin-lattice relaxation rates ($1/T_1^*$) for (a) TEMPO and (b) TEMPONE in protonated (●) and deuterated (○) ethanol glass. The experimental data were simulated (full line) according to Eq. (1) and the best-fit parameters are given in Table II.

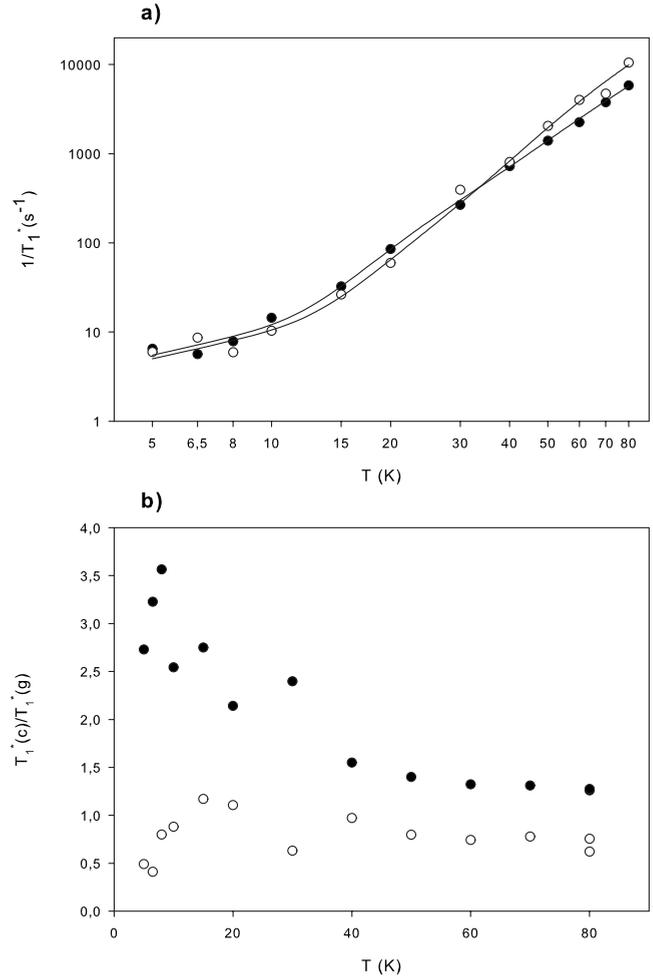

FIG. 2. (a) Temperature dependence of spin-lattice relaxation rates ($1/T_1^*$) for TEMPO (●) and TEMPONE (○) in crystalline ethanol. The experimental data were simulated (full line) according to Eq. (1) and the best-fit parameters are given in Table II. (b) The ratio of electron spin-lattice relaxation in crystalline ethanol, $T_1^*(c)$, and ethanol glass, $T_1^*(g)$, for TEMPO (●) and TEMPONE (○) as a function of temperature.

*Temperature dependence of* $1/T_1^*$. In order to analyze the underlying mechanisms responsible for different $T_1^*$ behavior of TEMPO and TEMPONE in the same state of the host matrix, the temperature dependence of spin-lattice relaxation rates was simulated as the sum of three contributions[16,21–23]

$$\frac{1}{T_1^*} = A\frac{T^9}{\theta_D^7}J_8(\theta_D/T) + Ce^{-\Delta/T} + BT \quad (1)$$

and the best-fit parameters are given in Table II. The first term of Eq. (1) is related to the two-phonon Raman process with $J_8$ denoting the transport integral and $\theta_D$ is the Debye temperature. The second term, presented in the low-temperature approximation form describes the low-frequency pseudolocal vibrations characterized by an energy, $\Delta$. It should be mentioned that in our previous study an estimation of motional correlation time of TEMPO at lowest temperatures studied was provided in the range of milliseconds.[5] This result was derived in the framework of slow-motional diffusion model[24] which we used for the analysis of phase memory time measurements. Since this type of dynamics is to slow to contribute to $T_1^*$ relaxation, in the context of this study we offer the explanation in terms of low-frequency pseudolocal vibrations which can arise due to dynamics of the guest molecule and/or can be related to low-frequency vibrational modes in the host matrix.[25,26] One can notice (Table II) that the energy of vibrations does not differ among the samples incorporating different guest molecules, pointing to the modes inherent to the ethanol host itself. The estimated energy of this mode is approximately 70 K, a result consistent with findings of neutron scattering, indicating a phonon band near 6 meV.[27]

TABLE II. The best-fit parameters in the simulation of the experimental spin-lattice relaxation rates according to Eq. (1). The values for Debye temperatures were taken from calorimetric measurements, $\Theta_D = 284/224$ K and $\Theta_D = 268/213$ K for protonated (H) and deuterated (D) ethanol in the crystalline (c)/glassy (g) state, respectively (Refs. 14 and 32).

| Sample | $B$ (K$^{-1}$ s$^{-1}$) | $C$ (s$^{-1}$) | $\Delta$ (K) | $A$ (s$^{-1}$ K$^{-2}$) |
|---|---|---|---|---|
| TEMPO (Hg) | $3.2 \pm 0.2$ | $2829 \pm 816$ | $64 \pm 6$ | $9 \pm 1$ |
| TEMPO (Dg) | $1.1 \pm 0.1$ | $3421 \pm 1680$ | $74 \pm 10$ | $7 \pm 2$ |
| TEMPONE (Hg) | $0.54 \pm 0.04$ | $1021 \pm 278$ | $58 \pm 5$ | $10.8 \pm 0.8$ |
| TEMPONE (Dg) | $0.65 \pm 0.06$ | $840 \pm 345$ | $52 \pm 5$ | $10 \pm 1$ |
| TEMPO (Hc) | $1.1 \pm 0.08$ | $3518 \pm 987$ | $81 \pm 7$ | $10 \pm 1$ |
| TEMPONE (Hc) | $1.0 \pm 0.1$ | $3424 \pm 2383$ | $88 \pm 16$ | $20 \pm 4$ |

The last term of Eq. (1) is important in the region of the lowest temperatures studied where the largest difference between $1/T_1^*$ data was detected (Table II). The linear temperature dependence of $1/T_1^*$ can be related to the presence of previously suggested excess of low-energy excitations in glasses that are not present in crystals.[28] The possible origin of these excitations is discussed as follows. (A) Localized quantum-mechanical two-level tunneling systems (TLS excitations) exist in glass but not in crystal and affect spin relaxation.[28] That this mechanism is important can be argued by the shorter $T_1^*$ of TEMPO in ethanol glass than crystalline ethanol. This reasoning is corroborated by the increased $T_1^*$ values of TEMPO incorporated in deuterated ethanol glass as compared with protonated one. Meanwhile, no change in $T_1^*$ for TEMPONE upon isotope substitution in the host matrix was observed along with the insensitivity to ethanol glass-crystalline transformation. Therefore, we can propose that, due to the hydrogen bonding between TEMPONE and ethanol, higher ordering in the local structure is induced. As a consequence, the density of TLS centers in close proximity to the spin probe is decreased and does not affect spin relaxation. Thus, a direct process can be assumed as a mechanism responsible for spin relaxation in both glassy and crystalline ethanol in the presence of TEMPONE. (B) The boson peak (BP) excitations refer to disordered solids and describe the excess in the vibrational density of states over that predicted by the Debye model and can contribute toward $1/T_1^*$ via a direct process.[29] In the context of EPR, BP excitations are expected to enhance the energy exchange between the spin system and the lattice. As a support for this reasoning, BP excitations, observed in Raman and neutron scattering experiments, are shown to contribute in ethanol specific heat maximally at approximately 6 K,[17] a result which can support our TEMPO $1/T_1^*$ data. The neutron-scattering approach revealed additionally that the BP intensity decreases as hydrogen-bond density increases in hydrogen-bonded molecular glasses.[30] This observation might explain why TEMPONE has a longer $T_1^*$ than TEMPO in ethanol glass, due to the lower density of low vibrational states and consequently the less efficient energy exchange between the spins and the lattice in the presence of the former rather than the latter paramagnetic dopant molecule. (C) The direct, one-phonon, process of acoustic phonons can contribute to spin relaxation in both crystalline and glassy matrices.[31] The intensity of the process scales with $\Theta_D^{-5}$ and the calorimetric measurements of solid ethanol[14,32] result in $\Theta_D^{glass} < \Theta_D^{crystal}$, which fits our experimental data appropriately.

The main difference between the TLS and BP mechanisms with respect to the direct process of acoustic phonons is that the former excitations exist only in glass while the latter ones could also be present in the crystal state. In addition, TLS and BP excitations are believed to be of local character as compared with the extended character of acoustic phonons. Since the greatest impact of ethanol matrix deuteration appears at the lowest temperatures studied, where the effect of order (crystal) versus disorder (glass) mostly affects $T_1^*$, this points to the importance of electron-nuclear dipolar interaction between the dopant molecule and the localized excitations (TLS, BP) in the host matrix. This mechanism should be responsible for the increase in TEMPO $1/T_1^*$ in ethanol glass versus crystalline ethanol. In this context, due to the hydrogen-bond network between TEMPONE and the host matrix, higher local ordering is established, causing a decrease in local excitations and rendering this guest molecule less sensitive to the dynamic constraints imposed by the local environment. For the detailed analysis of dominant excitations in ethanol glass versus crystalline ethanol, a multifrequency EPR study of spin-lattice relaxation should be considered.

This work was supported by the Croatian Ministry of Science, Education and Sports, Grants No. 098-0982915-2939 and No. 098-0982904-2912.

[1] J. I. Spielberg and E. Gelerinter, Phys. Rev. B **30**, 2319 (1984).
[2] G. Gradl and J. Friedrich, Phys. Rev. B **35**, 4915 (1987).
[3] T. Risse, W. L. Hubbell, J. M. Isas, and H. T. Haigler, Phys. Rev. Lett. **91**, 188101 (2003).
[4] S. N. Bhat, A. Sharma, and S. V. Bhat, Phys. Rev. Lett. **95**, 235702 (2005).
[5] M. Kveder, D. Merunka, M. Jokic, and B. Rakvin, Phys. Rev. B **77**, 094202 (2008).
[6] H. Sato, S. E. Bottle, J. P. Blinco, A. S. Micallel, G. R. Eaton, and S. S. Eaton, J. Magn. Reson. **191**, 66 (2008).
[7] Y. S. Bai and M. D. Fayer, Phys. Rev. B **39**, 11066 (1989).
[8] K. L. Ngai, J. Non-Cryst. Solids **353**, 709 (2007).
[9] D. A. Parshin, H. R. Schober, and V. L. Gurevich, Phys. Rev. B **76**, 064206 (2007).
[10] R. P. Wool, J. Polym. Sci., Part B: Polym. Phys. **46**, 2765 (2008).
[11] A. Zecevic, G. R. Eaton, S. S. Eaton, and M. Lindgren, Mol. Phys. **95**, 1255 (1998).
[12] A. Barbon, M. Brustolon, A. L. Maniero, M. Romanelli, and L. C. Brunel, Phys. Chem. Chem. Phys. **1**, 4015 (1999).
[13] G. Floridi, D. Brandis, O. Kanert, K. P. Dinse, and S. Cannistraro, Phys. Rev. B **48**, 13474 (1993).
[14] C. Talón, M. A. Ramos, S. Vieira, G. J. Cuello, F. J. Bermejo, A. Criado, M. L. Senent, S. M. Bennington, H. E. Fischer, and H. Schober, Phys. Rev. B **58**, 745 (1998).
[15] D. R. Allan and S. J. Clark, Phys. Rev. B **60**, 6328 (1999).
[16] M. Kveder, D. Merunka, A. Ilakovac, J. Makarević, M. Jokić, and B. Rakvin, Chem. Phys. Lett. **419**, 91 (2006).
[17] C. Talón, M. A. Ramos, and S. Vieira, Phys. Rev. B **66**, 012201 (2002).
[18] Arthur Schweiger and Gunnar Jeschke, *Principles of Pulse Electron Paramagnetic Resonance* (Oxford University Press, Oxford, 2001).
[19] B. Rakvin, N. Maltar-Strmečki, C. M. Ramsey, and N. S. Dalal, J. Chem. Phys. **120**, 6665 (2004).
[20] I. V. Blazhnov, S. Magazù, G. Maisano, N. P. Malomuzh, and F. Migliardo, Phys. Rev. E **73**, 031201 (2006).
[21] S. Lijewski, M. Wencka, S. K. Hoffmann, M. Kempinski, W. Kempinski, and M. Sliwinska-Bartkowiak, Phys. Rev. B **77**, 014304 (2008).
[22] P. G. Klemens, Phys. Rev. **125**, 1795 (1962).
[23] J. Murphy, Phys. Rev. **145**, 241 (1966).
[24] S. Lee and S. Z. Tang, Phys. Rev. B **31**, 1308 (1985).
[25] J. Goslar, S. K. Hoffman, and W. Hilczer, Solid State Commun. **121**, 423 (2002).
[26] Yu. G. Vainer, A. V. Naumov, and L. Kador, Phys. Rev. B **77**, 224202 (2008).
[27] M. A. Ramos, S. Vieira, F. J. Bermejo, J. Dawidowski, H. E. Fischer, H. Schober, M. A. González, C. K. Loong, and D. L. Price, Phys. Rev. Lett. **78**, 82 (1997).
[28] M. K. Bowman and L. Kevan, J. Phys. Chem. **81**, 456 (1977).
[29] N. P. Giorgadze and L. Zh. Zakharov, Low Temp. Phys. **24**, 198 (1998).
[30] O. Yamamuro, K. Takeda, I. Tsukushi, and T. Matsuo, Physica B **311**, 84 (2002).
[31] J. H. Van Vleck, Phys. Rev. **57**, 426 (1940).
[32] M. A. Ramos, C. Talón, and S. Vieira, J. Non-Cryst. Solids **307-310**, 80 (2002).